\documentclass[10pt, aps, prd,
reprint, % for double column
notitlepage, nofootinbib, longbibliography,
superscriptaddress,
]{revtex4-1}

\usepackage{amsmath,bm,amsfonts}
\usepackage{pstricks,pstricks-add}
\usepackage{pst-all,pst-slpe}

\usepackage{graphicx}
\usepackage[breaklinks=true,colorlinks=true,urlcolor=blue,linkcolor=red,
    citecolor=red]{hyperref}

\begin{document}

\title{Geometrical dependence in Casimir-Polder repulsion}
%\title{Geometrical dependence of Casimir-Polder repulsion for weak aperatured materials}

\author{John Joseph Marchetta}
\email{jjmarchetta@gmail.com}
\affiliation{Department of Physics,
Southern Illinois University--Carbondale, Carbondale, Illinois 62901, USA}

\author{Prachi Parashar}
\email{Prachi.Parashar@jalc.edu}
\affiliation{John A. Logan College, Carterville, Illinois 62918, USA}

\author{K. V. Shajesh}
\email{kvshajesh@gmail.com}
\affiliation{Department of Physics,
Southern Illinois University--Carbondale, Carbondale, Illinois 62901, USA}

\date{\today}
%--------------------------------------------

\begin{abstract}
Repulsion, induced from quantum vacuum fluctuations, for an anisotropically
polarizable atom on the symmetry axis of an anisotropically polarizable
annular disc is studied. There exists two torsion free points on each side
of the annular disc, where the interaction energy is orientation
independent. The position of second of the two torsion free points, on
either side of the disc, is shown to determine the orientation dependence
of the atom at distances far from the plate, revealing a geometrical
dependence. In the ring limit of the annular disc, new repulsion emerges.
\end{abstract}

\maketitle
%--------------------------------------------

%------------------------

The Casimir effect is an umbrella term referring to interactions
induced by quantum fluctuations in the electromagnetic field. The
associated forces become dominant in the nanoscale, for neutral
configurations. The related van der Waals and London dispersion forces
have to do with the domain in which the materials interact weakly at short
distances~\cite{Waals:1873sl,London:1930a,London:1930b,Hettema:2001cq}.
Casimir and Polder~\cite{Casmir:1947hx} extended London dispersion force
to include retardation effects at large distances using fourth order
perturbation theory. Casimir then computed the attraction between two
neutral conducting plates, which was inaccessible to perturbation theory,
by evaluating the electromagnetic zero point energy outside and inside
the plates with only the difference contributing to a finite inward
pressure~\cite{Casimir:1948pc}. Thus, Casimir's result suggested
that the zero point energy could have measurable consequences.

Repulsive forces from the Casimir effect counter the intuition given
for its usual attractive nature. Nevertheless, the first repulsive
forces were found early on in three-body van der Waals interactions by
Axilrod and Teller~\cite{Axilrod:1943at} and Muto~\cite{Muto:1943fc},
and by Craig and Power~\cite{Craig1969ma,Craig:1969pa}
in Casimir-Polder interactions between anisotropically
polarizable atoms. The counterintuitive nature of repulsion induced
from quantum fluctuations explains why it
was not until a decade ago that a repulsive effect was
realized between conductors in the (strong and retarded)
nonperturbative Casimir regime.
This came by in Ref.~\cite{Levin:2010vo}, where it was
argued on physical grounds that the interaction energy
between an elongated conductor and a perfectly conducting
metal sheet with a circular aperture could have a local minima.
They showed that the Casimir force could become repulsive
when the needle got sufficiently close to the aperture.
It was suggested that the anisotropy in the geometry of a highly
conducting object corresponds to an effective anisotropic permittivity
and the interplay of these anisotropic permittivities
leads to non-monotonic interaction energies that
causes repulsion~\cite{McCauley:2011rd,Shajesh:2011daa}. 
Only few attempts have been made to understand
the result in Ref.~\cite{Levin:2010vo} analytically.
In Ref.\cite{Eberlein2011hwp}, 
using the method of inversion from electrostatics,
it was shown that even in the van der Waals
(weak and non-retarded) limit,
similar configurations lead to repulsion.
In Ref.~\cite{Abrantes2018tcp}, the method of inversion was again
used to show that in the (weak and non-retarded) van der Waals regime
an anisotropically polarizable atom placed along the 
symmetry axis of a toroid could experience repulsion. 
An advance was made in the retarded regime when
an analytic formula for the (weak and retarded)
Casimir-Polder energy between
dielectric bodies was derived in
Refs.~\cite{Shajesh:2011daa} and \cite{Milton2012esc},
demonstrating that repulsion is possible in
configurations with anisotropically polarizable atoms
and anisotropic dielectric materials.
In spite of the above successes in the weak scenario,
analytic derivation in the strong regime of a closed-form expression for
the interaction energy between a polarizable atom
and a highly conducting plate with an aperture still remains 
elusive~\cite{Milton2011asa,2012:Miltonfpc,Shajesh2017ssa}.

Repulsion in Casimir effect also appears if the interacting
bodies possess certain electric and magnetic
properties~\cite{Boyer:1974van}, or if the medium's conductivity
is greater than that of the materials'~\cite{Dzyaloshinskii:1961fw}.
However, in the current level of our understanding, apparently,
these repulsions have different origins from those discussed in
this paper: that of interaction energies between
anisotropically polarizable atoms and anisotropically
polarizable dielectrics.

%-------
\begin{figure}
\includegraphics[width=3cm]{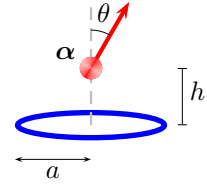}
\caption{Point atom of polarizability ${\bm\alpha}$
above a dielectric ring of polarizability ${\bm\sigma}$.
The atom is on the symmetry axis of the ring. }
\label{point-atom-above-ring-fig}
\end{figure}%
%-------

In this Letter, we analyze the interaction,
including the orientation dependence, of
an anisotropically polarizable atom above the center of an
anisotropically polarizable ring 
depicted in Fig.~\ref{point-atom-above-ring-fig}
using the methods of Refs.~\cite{Shajesh:2011daa,Milton2012esc}.
A full exposition of the interaction for each polarization basis
of the ring and the atom is available in Ref.~\cite{Marchetta:2020dap},
which serves as supplementary material for this article. 
Casimir-Polder repulsion for an atom interacting with the
ring demonstrates a richer dependence on polarizability than the
corresponding interaction with a plate with a circular aperture in
Ref.~\cite{Shajesh:2011daa}.
In Refs.~\cite{Shajesh:2011daa,Milton2012esc}
it was shown that the atom gets repelled at short distances
only when its orientation deviates no more than a critical angle 
away from the axis perpendicular to the plate. An orientation
independent point, where the energy is independent of the orientation
of the atom and thus feels no torsion there, was found on both sides
of the plate. It was established that the energy is minimized
when its orientation is parallel to that of the plate for distances
less than the torsion free point, and perpendicular to the plate
for larger distances. 

\begin{figure}
\includegraphics[width=8cm]{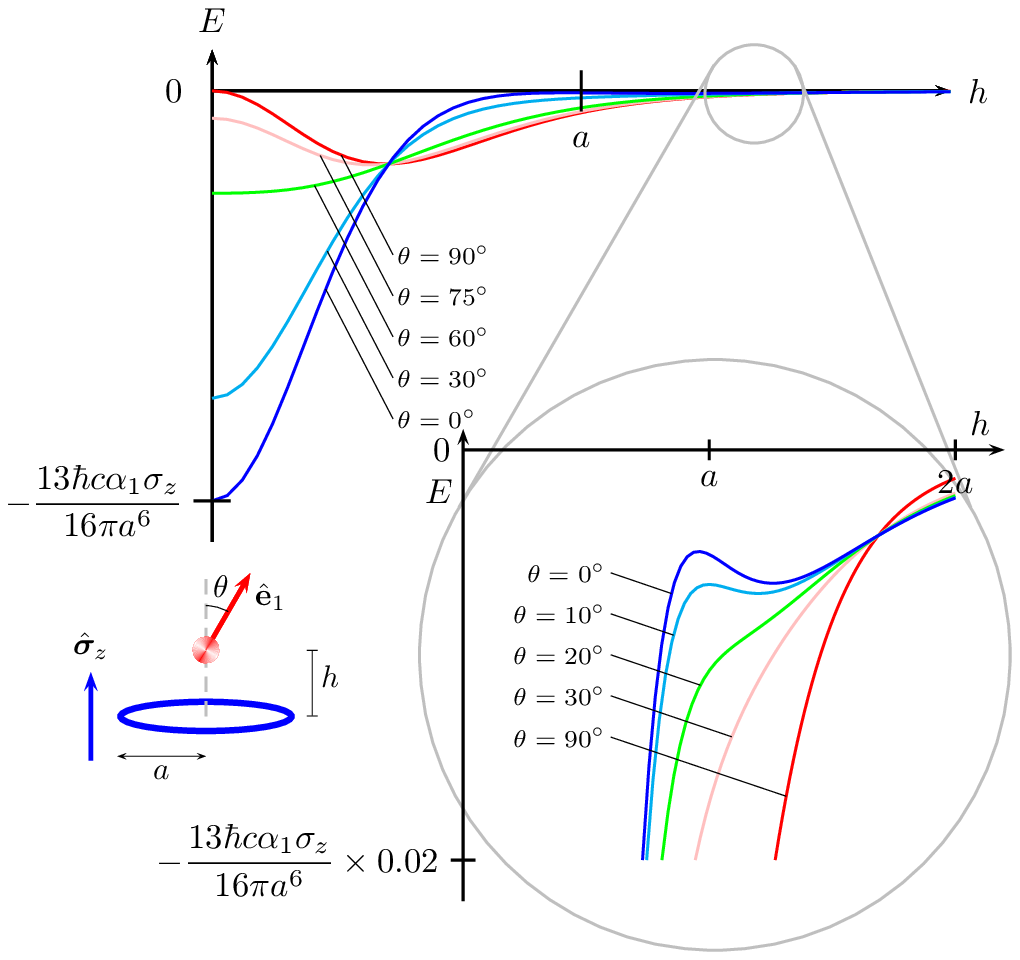}
\caption{The interaction energy between an atom with polarizability
${\bm\alpha} =\alpha_1\, \hat{\bf e}_1 \hat{\bf e}_1$
and a ring of polarizability
${\bm\sigma} = \sigma_{z}\, \hat{\bm z} \hat{\bm z}$
is illustrated in bottom left corner of figure.
The energy in Eq.\,(\ref{intE-e1lr}) is plotted as function of
height $h$ for different orientations of $\hat{\bf e}_1$
with respect to the symmetry axis of the ring.
The first region of repulsion occurs for $|h|<0.47\,a$ 
and $60.88^\circ <\theta <119.12^\circ$. 
A second region of repulsion is not visible, however,
the inset provides a zoomed in view of this non-monotonicity,
where the switch from attraction to repulsion occurs.
Repulsive force experienced by the atom in the direction of
the symmetry axis will be a manifestation of negative slopes
in these plots.}
\label{atom-above-dielectric-ringe1Lr-fig}%
\end{figure}

We show that, for certain polarizabilities of the ring
in Fig.~\ref{point-atom-above-ring-fig},
see Ref.~\cite{Marchetta:2020dap} for  details,
there exist two torsion-free heights on each side of the ring,
in contrast to a single torsion free point on each side for a
plate with circular aperture.
We show that for specific orientations the atom could feel
a repulsive force when it is near the center of the ring,
and, in addition, at intermediate distances from the center
of the ring,
for example see Fig.~\ref{atom-above-dielectric-ringe1Lr-fig}.
To understand this additional repulsive region, and the additional
torsion-free point, we analyze
the configuration involving an atom interacting with an
anisotropically polarizable annular disc.
Keeping the inner radius of the annular disc fixed and varying the
outer radius, we can smoothly deform the disc into a ring, or
a plate of infinite extent with a circular aperture.
We find upper bounds on the outer radius of the disc that allows 
for repulsion at intermediate distances.
The second torsion-free height moves farther away as the outer radius
of the disc increases. Interestingly, even though one torsion-free point
moves to infinity in the plate limit for certain polarizabilities,
it does not move afar indefinitely for all the polarizabilities.

In this Letter we focus our attention to the case when the ring
is polarizable in the direction of the symmetry axis of the ring.
For an extensive analysis we refer to Ref.~\cite{Marchetta:2020dap}.
The discussion in this article and repulsive Casimir effect in general
is expected to be of considerable interest in nanotechnology,
because it could assist in reducing `stiction' among the 
components within nano devices. As a simpleminded application of
a configuration studied here, we entertain the plausibility that
the interaction between a ring and atom could serve as a prototype
for a rotaxane. We discuss an idealized cycle that could be
implemented when designing such molecular machines.  

The Casimir-Polder interaction energy between dilute dielectric bodies
is given in terms of the electric susceptibilities
${\bm\chi}_i$, $i=1,2,$ of the bodies as~\cite{Marchetta:2020dap}
\begin{equation}
E_{12} = -\frac{\hbar c}{2} \int_{-\infty}^{\infty}
\frac{d\zeta}{2\pi}\,\text{Tr} \Big[ 
{\bf \Gamma}_0\cdot {\bm \chi}_1 \cdot {\bf \Gamma}_0
\cdot {\bm \chi}_2 \Big],
\label{E12=Logof}
\end{equation}
where the trace $\text{Tr}$ is over both the coordinate and
matrix indices.
The polarizable atom in Fig.~\ref{point-atom-above-ring-fig}
is suitably described in terms of the polarizability tensor
${\bm\alpha}(\omega)$ of the atom, 
${\bm\chi}_1({\bf r})
={\bm\alpha}(\omega)\delta^{(3)}({\bf r}-{\bf r}_0)$,
where ${\bf r}_0$ is the position of the atom.
The ring, similarly, is described in terms of the polarizability
per unit length ${\bm\sigma}(\omega)$ of the ring,
${\bm\chi}_2({\bf r})={\bm\sigma}(\omega)\delta(\rho-a)\delta(z)$,
where $a$ is the radius of the ring. 
The polarizabilities are dependent on frequency $\omega$ and
the interaction energy is expressed as an integral over the 
imaginary frequency $\omega =i\zeta$. The free Green dyadic
${\bm\Gamma}_0$ in Eq.\,(\ref{E12=Logof}) is given by
\begin{equation}
{\bf \Gamma}_0({\bf r};i\zeta) = \frac{e^{-|\zeta|r}}{4\pi\,r^3}
\Big[ -u(|\zeta|r) \,{\bf 1} + v(|\zeta| r)\, \hat{\bf r} \hat{\bf r} \Big]
\end{equation}
with $u(x)=1+x+x^2$ and $v(x)=3+3x+x^2$.
For large distances between the bodies, the interaction energy
in Eq.\,(\ref{E12=Logof}) only gets contributions
from very small frequencies. In the Casimir-Polder limit,
the polarizabilities can be approximated by their static
values, $\omega=i \zeta=0$, and then the $\zeta$-integration
can be completed easily. 

Here we consider
${\bm \alpha}(0)=\alpha_{1}\hat{\bf e}_{1}\hat{\bf e}_{1}$
and ${\bm \sigma}(0)=\sigma_{z}\hat{\bf z}\hat{\bf z}$,
where $\hat{\bf z}$ is chosen to be in the direction of the symmetry
axis of the ring and $\hat{\bf e}_1$ represents the orientation
of the polarizability of the atom as described in
Fig.~\ref{point-atom-above-ring-fig}.
The interaction energy for the atom and ring is found 
to be~\cite{Marchetta:2020dap}
\begin{equation}
\begin{aligned}
E = -\frac{\hbar c\alpha_1\sigma_z}{64\pi}
\frac{a}{(a^2+h^2)^\frac{11}{2}}
\Big[ (26 a^4 + 3h^2 a^2 + 40 h^4)&\\
+ (26 a^4 - 123 h^2 a^2 + 40 h^4) \cos 2\theta \Big],
\label{intE-e1lr}
\end{aligned}
\end{equation}
where the angle $\theta$ is the deviation of the atom's orientation
from the symmetry axis of the ring in
Fig.~\ref{point-atom-above-ring-fig},
while $h$ and $a$ denote the height of the atom above the ring
and the radius of the ring, respectively.
%------
%------
In Fig.~\ref{atom-above-dielectric-ringe1Lr-fig}, we plot the 
interaction energy of Eq.\,(\ref{intE-e1lr}) as a function of
height $h$ for various orientations $\theta$. The energy is
minimized at $h=0$ for $\theta=0^\circ$, suggesting that for
$h=0$ the atom tries to align its direction of
polarizability with that of the ring.
The heights for which the interaction energy is orientation
independent and torsion-free is determined when the coefficient of
$\cos(2\theta)$ in Eq.\,(\ref{intE-e1lr})
is zero, which is at ${h}=\pm 0.48\,a$ and $h=\pm\,1.69a$.
Upon crossing a torsion-free point, the orientation that minimizes
the energy changes by $90^{\circ}$ as shown in the inset
of Fig~\ref{atom-above-dielectric-ringe1Lr-fig}.
Repulsive forces on the atom requires the parameters
to satisfy the inequality
\begin{equation}
\frac{h}{a} < \pm \sqrt{\frac{(19+181\cos{2\theta})\pm
\sqrt{D}}{80(1+\cos{2\theta})}},
\label{rep-reg-zz-ex}
\end{equation}
where
\begin{equation}
D= -6039 -11682 \cos 2\theta +20601\cos^2 2\theta.
\label{dis-tinqeLz}
\end{equation}
Repulsion occurs when $60.88^\circ<\theta<119.12^\circ$
for short distances and has the maximum extent in height
for $|h|< 0.47\,a$ at $\theta=90^{\circ}$.
When $|\theta|<13.27^\circ$,
repulsion emerges for a range of intermediate distances.
This range is of maximum extent when $\theta=0^\circ$, for 
$1.24\,a<|h|< 1.41\,a$. The repulsive force at these
intermediate distances, though small in magnitude,
might lead to interesting applications.

We consider an annular disk to understand the
differences between the ring and an aperatured plate.
The annular disc is described by 
${\bm\chi}_{2}({\bf r})
={\bm\lambda}(\omega)\theta(b-\rho)\theta(\rho-a)\delta(z)$,
where ${\bm\lambda}(\omega)$ is the polarizability per unit
area of the annular disc. Here $b$ is the outer radius of
the disc and $a$ is the inner radius.
We consider the case 
${\bm \lambda}(0)=\lambda_{z}\hat{\bf z}\hat{\bf z}$.
The limit as $b\to \infty$ 
corresponds to a plate with a circular aperture of radius $a$.
Taking the delicate limit of $b\to a$ with
the surface polarizability $\lambda_{z}\to\infty$,
such that $\sigma_{z}=\lambda(b-a)$ is kept fixed,
constructs a ring~\cite{Marchetta:2020dap}.
The energy of the annular disc is computed to be~\cite{Marchetta:2020dap}
\begin{eqnarray}
E &=& -\frac{\hbar c\alpha_1\lambda_z}{64\pi} \frac{1}{5}
 \frac{(-1)}{(\rho^2+h^2)^\frac{9}{2}}
\Big[ (26 \rho^4 + 17h^2 \rho^2 + 26 h^4)
\nonumber \\ && \hspace{10mm}
+ (26 \rho^4 -73 h^2 \rho^2 +6 h^4) \cos 2\theta \Big]
\bigg|^{\rho=b}_{\rho=a}.
\label{CPannD-e1lz-apr}
\end{eqnarray}
Unlike the case of ring, in Eq.\,(\ref{rep-reg-zz-ex}),
the repulsive regions can not be captured in a closed-form 
expression for the case of a disc, but can be easily determined
numerically. The intermediate region of repulsion,
as described in the zoomed in view of 
Fig.~\ref{atom-above-dielectric-ringe1Lr-fig} for a ring,
vanishes when the width of the annulus is above the critical values
given in Table~\ref{table-outer-rad} for the specific orientations
of the atom.
%-------
\begin{table}
\begin{tabular}{cc} 
\hline
Orientation & \hspace{6mm} Criteria for second region of repulsion \\ [0.5 ex] 
\hline\hline
$\theta=0^{\circ}$ & $a<b<1.6505a$ \\ \hline
$\theta=6^{\circ}$ & $a<b<1.559a$\\ \hline
$\theta=9^{\circ}$ & $a<b<1.4379a$ \\ \hline
$\theta=12^{\circ}$& $a<b<1.2323a$ \\ \hline
\end{tabular}
%-------
\caption{This table lists criteria for the region of repulsion
for intermediate distances repulsion to occur between an
anisotropically polarizable atom and an annular disc with
polarizability in the $z$-direction. The second column
lists the upper bound on the outer radius $b$ of the disc
for orientation angles $\theta$ of the atom. }
\label{table-outer-rad}
\end{table}
The four orientation independent heights,
$\pm h_1$ and $\pm h_2$, for the case of disc are determined by
\begin{equation}
\frac{(26b^4 -73b^2h^2+6h^4)}{(b^2+h^2)^\frac{9}{2}}
-\frac{(26a^4 -73a^2h^2+6h^4)}{(a^2+h^2)^\frac{9}{2}} =0,
\end{equation}
with the solutions above the disc satisfying
\begin{subequations}
\begin{equation}
0.48 a\xleftarrow[\text{ring}]{a\leftarrow b}
|h_1| \xrightarrow[\text{plate}]{b\to\infty} 0.60 a
\end{equation}
and
\begin{equation}
1.69 a\xleftarrow[\text{ring}]{a\leftarrow b}
|h_2| \xrightarrow[\text{plate}]{b\to\infty} 3.44a.
\end{equation}
\end{subequations}

A difference in the systematics of the features in the energy
occurs between axial versus radial polarizability of the 
disc~\cite{Marchetta:2020dap}.
The four orientation independent heights for the case of
radial polarizability of the disc are given by
\begin{subequations}
\begin{equation}
0.36 a\xleftarrow[\text{ring}]{a\leftarrow b} 
|h_1| \xrightarrow[\text{plate}]{b\to\infty} 0.44 a
\end{equation}
and
\begin{equation}
3.45 a\xleftarrow[\text{ring}]{a\leftarrow b}
|h_2| \xrightarrow[\text{plate}]{b\to\infty} \infty.
\end{equation}%
\label{thrho}%
\end{subequations}%
Unlike the case of radial polarizability above,
in the case of axial polarizability the two orientation
independent points, two on each side of the disc,
exist even in the limit when the disc
becomes a plate. In this scenario,
the second orientation independent height increases to a
finite value in the plate limit.
This implies that the orientation preference of the atom at
$h=0$ and $h\to\infty$ is the same for axial polarizability.
This should be contrasted with that of radially
polarizable disc in which case the orientation preference
of the polarizability of the atom
at $h=0$ is orthogonal to that at $h\to\infty$.

%--------
\begin{figure}
\includegraphics[width=8.5cm]{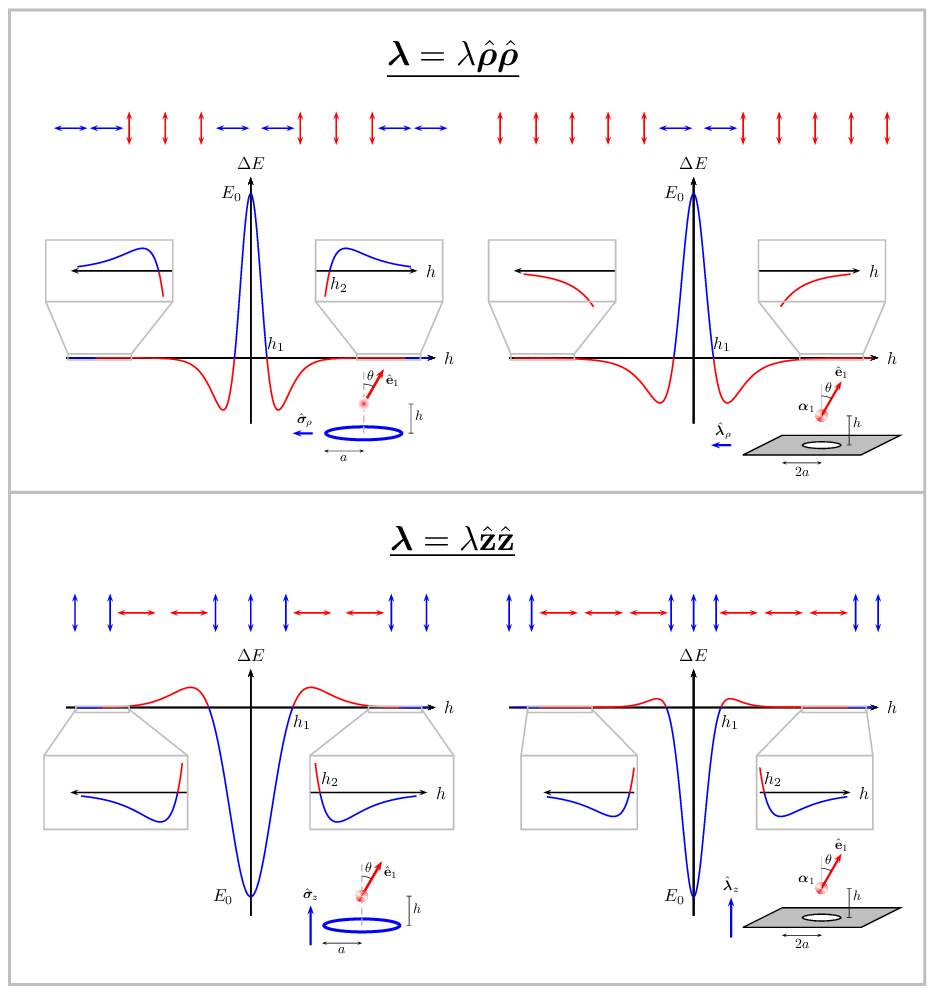}
\caption{Difference in interaction energy $\Delta E$ between
$\theta=0$ and $\theta=\pi/2$ plotted as a function of height $h$.
The intersection of the energy curves with abscissa 
represent orientation independent points, marked 
$h_1$ and $h_2$. The orientation dependence at each height is shown
as double-arrowed vectors adjacent to the horizontal axis.
Observe that the orientation preferences for the plate
is different for radial and axial polarizabilities.
}
\label{E-versus-h-summary-fig}
\end{figure}%
%-------

We summarize our results in Fig.~\ref{E-versus-h-summary-fig},
which highlights the orientation independent torsion free points
and the domains of repulsion.
For large distances from the aperture the atom tends
to orient its polarizability perpendicular to the
polarizability of the plate when the plate is polarizable
in the radial direction, while it tends to orient parallel
when the plate is polarizable in the axial direction. This
feature at $h\to\infty$ is completely determined by the
the existence of the second torsion free point.
The repulsion felt at intermediate distances is about six 
orders of magnitude weaker than the repulsion at short distances 
which still lacks experimental evidence. It is not clear
if the repulsions found for intermediate region of heights 
will persist when the conductivity of the plate is larger.

%--------
\begin{figure}
\includegraphics[width=8cm]{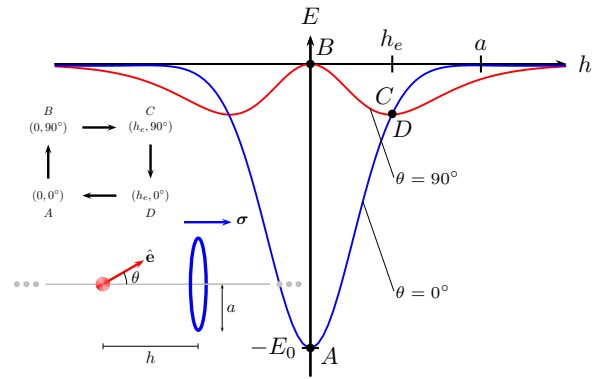}
\caption{The bottom left part of the figure displays a prototype
of a Casimir machine using an atom and a ring. The interaction energy
of the configuration as a function of height $h$ for $\theta=0$
and $\theta=90^{\circ}$ are plotted. The points 
$A$, $B$, $C$, and $D$, on the plots are used to construct
a cyclic path. }
\label{casimir-machine-fig}
\end{figure}
%--------

As a simpleminded application of the configuration discussed here,
we propose a prototype of Casimir machine along the lines of
a rotaxane~\cite{Anelli:1991sma}, rendering its applicability
to molecular transportation, molecular switching, and nanorecording. 
By requiring the atom to be fixed on an axis and allowing the ring
to move with center on the axis, we consider a cyclic path involving
the points $A\to B\to C\to D \to A$
in Fig.~\ref{casimir-machine-fig}. The atom starts at the center
of the ring while orientated parallel to the polarizability of the ring,
denoted by $A$. We put work into the system to rotate the atom
by $90^{\circ}$ so that the energy is maximized at point $B$.
The ring is then repelled from the atom to $h=0.47a$.
By expending a small amount of energy
the ring adjusts its position to the torsion free point at
$h_{e}=0.48a$ labeled in the diagram by $C$. Position $D$
represents the ring at $h_{e}$ but with the atom rotated,
without cost of internal energy, to $\theta=0$.
It is energetically favorable for the atom to then
return to the center of ring and complete the cycle.

In summary, the interaction between an anisotropically polarizable
dielectric annulus and an anisotropically polarizable atom
reveals tight constraints on their relative geometrical orientations
in their directions of polarizability for repulsion. 
We have shown that the interaction energies are characterized
by torsion free points and domains of repulsion. The intuition
gained from this study will guide us in finding a complete
understanding of Casimir repulsion.
 
%-----------------------------------------------------
%\acknowledgments

JJM acknowledges support from the REACH Program
at Southern Illinois University--Carbondale.
PP and KVS remember Martin Schaden for collaborative assistance.

%-----------------------------------------------------
%\bibliographystyle{unsrt}
\bibliography{biblio/b20131003-casimir-top}
%\nocite{*} %%% Will print the complete bib data.
%-----------------------------------------------------

\end{document}